\documentclass[useAMS,usenatbib]{mn2e}
\usepackage{aas_macros}
\usepackage{graphicx,amssymb,natbib,textcomp,fontenc,lscape,color,afterpage}
\usepackage{deluxetable}
\newcommand{\ngc}{{\small NGC\,7469}}
\newcommand{\carma}{{\small CARMA}}
\newcommand{\iram}{{\small IRAM}}
\newcommand{\atca}{{\small ATCA}}
\newcommand{\vla}{{\small JVLA}}
\newcommand{\vlbi}{{\small VLBI}}
\newcommand{\merlin}{{\small MERLIN}}
\newcommand{\alma}{{\small ALMA}}
\newcommand{\xmm}{{\small \it XMM-Newton}}
\newcommand{\swift}{{\small \it Swift}}

\newcommand{\xrt}{{\small {\it Swift}/XRT}}
\newcommand{\uvot}{{\small {\it Swift}/UVOT}}
\newcommand{\fv}{F_\mathrm{var}}
\newcommand{\Lnu}{erg\,s$^{-1}$Hz$^{-1}$}

\title[Millimeter-wave and X-ray monitoring of NGC\,7469]
{Simultaneous Millimeter-wave and X-ray monitoring of the Seyfert Galaxy NGC\,7469}
\author[Behar et al.]
{Ehud Behar$^1$\thanks{E-mail: behar@physics.technion.ac.il},  
Shai Kaspi$^2$, 
Gabriel Paubert$^3$, 
Nicolas Billot$^4$,
Uria Peretz$^1$
\newauthor Ranieri D. Baldi$^{5,6}$,
Ari Laor$^1$, 
Jelle Kaastra$^7$,
Missagh Mehdipour$^7$
\\
$^1$Department of Physics, Technion 32000, Haifa 32000, Israel\\
$^2$Wise Observatory and School of Physics and Astronomy, Tel Aviv University, Tel Aviv 69978, Israel\\
$^3$Instituto de Radioastronom'a MilimŽtrica, Avda. Divina Pastora 7, Nœcleo Central, E-18012 Granada, Spain\\
$^4$Observatoire Astronomique de l'universit{\'e} de Gen{\`e}ve, 51 Chemin des Maillettes, 1290 Versoix, Switzerland\\
$^5$School of Physics and Astronomy, University of Southampton, UK\\
$^6$Dipartimento di Fisica, Universitˆ degli Studi di Torino, via Pietro Giuria 1, I-10125 Torino, Italy\\
$^7$SRON Netherlands Institute for Space Research, Sorbonnelaan 2, 3584 CA Utrecht, The Netherlands
}
\begin{document}



\maketitle

\label{firstpage}

\begin{abstract}

We report on daily monitoring of the Seyfert galaxy \ngc , around 95\,GHz and 143\,GHz, with the \iram\ 30\,m radio telescope, and with the \swift\ X-Ray and UV/Optical telescopes, over an overlapping period of 45 days.
The source was observed on 36 days with \iram , and the flux density in both mm bands was on average $\sim 10$\,mJy, but varied by $\pm50\%$, and by up to a factor of 2 between days.
The present \iram\ variability parameters are consistent with earlier \carma\ monitoring, which had only 18 data points.
The X-ray light curve of \ngc\ over the same period spans a factor of 5 in flux with small uncertainties. 
Similar variability in the mm-band and in the X-rays lends support to the notion of both sources originating in the same physical component of the AGN, likely the accretion disk corona.
Simultaneous monitoring in eight UV/optical bands shows much less variability than the mm and X-rays, implying this light originates from a different AGN component, likely the accretion disk itself.
We use a tentative 14\,day lag of the X-ray light curve with respect to the 95\,GHz light curve to speculate on coronal implications.
More precise mm-band measurements of a sample of X-ray-variable AGN are needed, preferably also on time scales of less than a day where X-rays vary dramatically, in order to properly test the physical connection between the two bands.

\end{abstract}


\section{Introduction}

Radio loud (RL) active galactic nuclei (AGN) are known to have relativistic jets on various scales, from pc to Mpc.
On the other hand, radio emission from radio quiet (RQ) AGN is several orders of magnitude weaker, and has several possible origins. The implied high brightness temperature of these sources strongly suggests the presence of hot, non-thermal electrons,
but the actual physical origin and location of the radio source in RQ AGN is still in debate.
The nature of radio emission from RQ AGN could be due to weak jets, nuclear star formation, an AGN wind, or accretion-disk magnetic (i.e., coronal) activity.
For a recent review see \citet{Panessa19}.  
Some \vlbi\ images \citep{Anderson04, Ho01, Lal04, ulvestad05_rqq} indicate source angular sizes as small as mas, or pc-scales for nearby sources, which can still be consistent with all of the above origins.
Since these radio sources could be smaller than our contemporary imaging capability, one needs to revert to spectroscopy and timing in the attempt to characterize their physical origin.

\subsection{Spectra}
The radio spectra of RQ AGN have not been explored much.
Recently, \citet{Laor19} studied the spectral slope of 25 PG quasars \citep{schmidt83} between 5 and 8.5\,GHz, finding that a steep spectral slope of the flux density $F_\nu \propto \nu^\alpha$ with $\alpha < -0.5$ occurs in highly accreting quasars with $L/L_\mathrm{Edd} > 0.3$, while flatter slopes of  $\alpha > -0.5$ occur in quasars with $L/L_\mathrm{Edd} < 0.3$. 
The steep slope is typical of optically-thin synchrotron and was associated with an AGN wind, while a flat spectrum is indicative of an optically-thick source, and was associated with a compact AGN core.
More generally, since the sources are non-thermal and the emission is likely due to synchrotron, observing at a few GHz limits the size of an opaque synchrotron source one has a chance to resolve, either angularly, or temporally.
Synchrotron self-absorption at a few GHz makes a source of $L_ \nu \sim 10^{30}$ \Lnu\ opaque at the pc scales \citep[e.g., Eq. 19 in][]{laor08}.
Hence, smaller scales are physically impossible to observe.
Luckily, the synchrotron absorption coefficient $\alpha _\nu$ decreases strongly with frequency ($\propto \nu ^{-3}$), 
and thus the physical size of an opaque source decreases as well  \citep[$\propto \nu^{-5/4}$,][]{laor08}.
Consequently, observations at increasingly higher frequencies have the potential to reveal smaller structures close-in to the center of the AGN.
This effect is further enhanced by the improved angular resolution with $\nu$ of the telescopes.

Above 300\,GHz (sub-mm), the AGN spectrum rises steeply due to thermal dust emission \citep{barvainis92, hughes93}.
Thus, the band between 30--300~GHz (crudely referred to here as the mm band) is the sweet spot in which one might be able to observe nuclear activity.
This potential of observations around 100\,GHz has led to several mm-wave campaigns of RQ AGN.
\citet{doi05, doi11} observed a sample of low luminosity AGN and early type galaxies, both RQ and RL.
100\,GHz (3\,mm) fluxes of X-ray selected RQ AGN measured with \atca\ and \carma\ \citep{Behar15, Behar18} indicate a spectral turnover between the cm and the mm bands.
This distinct mm-wave spectral component appears superimposed on a steep spectrum of the more extended (optically thin) low-frequency source \citep[perhaps an outflow? see][]{Laor19}.
If this new spectral component is due to optically thick synchrotron, it could be the sign of a much smaller source than the likely extended few-GHz source.
More comprehensive cm- to mm- wave spectra were recently obtained by \citet{Inoue18}, who used \vla\ and \alma\ to clearly identify this component in NGC\,985 and in IC\,4329A.
These authors modeled their spectra in terms of coronal magnetic activity arising only a few 10s of Schwarzschild radii from the central black hole. 
Earlier theoretical works have explored the viability of coronal emission from around and near the black bole \citep{Field93, Inoue14, Raginski16}.

\subsection{Variability}
Weak variability at 8.5\,GHz on time scales of a few days for low-luminosity RQ AGN \citep{Anderson05}, and of months for luminous RQ quasars \citep{Barvainis05} has been detected.
Variability of RQ AGN typically $<$50\% at 5-15 GHz on a year time scale has also been detected \citep{neff83,wrobel00,mundell09}.
\citet{falcke01} found that RQ AGN are more variable than RL AGN except for blazars, and that low-luminosity RQ AGN are among the most variable radio sources, likely due to the small black-hole mass and size.

Systematic radio monitoring of RQ AGN cores is difficult, both due to the numerous observations required, and due to the faintness ($\sim$mJy) of the sources, which necessitates large interferometers, or few-hour exposures with smaller telescopes each time, as we do here.
The size of an optically thick self-absorbed synchrotron source at a few GHz being around a pc, further underlines the need for high-frequency (mm-wave) observations, and in particular monitoring.

\citet{doi11} found significant 3-mm variability within months in a few low-luminosity RQ AGN.
The only long-term mm-wave monitoring of a RQ AGN is that of \citet{baldi15}, who monitored \ngc\ with \carma\ at 95 GHz over a period of 70 days. The source varied ($\pm 3\sigma$ from the mean) by a factor of 2 within 4-5 days, unlike its long-term steadiness at 8.4\,GHz \citep{pereztorres09}.

\subsection{Connection to X-rays}
X-rays in RQ AGN vary rapidly, particularly in Seyfert galaxies on time scales of hours and less \citep{markowitz04}, 
which indicates they come from small length scales (the corona, $\lesssim$ light-hours), close to the black hole.
If the mm-wave emission comes from the same hot coronal electrons that produce the X-rays, the two bands would be physically correlated.
In terms of luminosity, \citet{laor08} used the PG quasar sample \citep{schmidt83} to demonstrate that $L_R$ and $L_X$ are correlated and follow the well established $L_R/L_X\sim 10^{-5}$ correlation observed in coronally active cool stars \citep{gudel93}.
In terms of variability, the sole mm-wave monitoring campaign of \citet{baldi15} indicate the variability parameters of \ngc\ at 95\,GHz are similar to those of \ngc\ in (non-simultaneous) archival X-ray data, only with much larger uncertainties.
However, in order to truly test the connection between the mm-wave and X-ray sources of RQ AGN, simultaneous monitoring is essential.
Diligent attempts at 5 and 8.5\,GHz, either focus on more RL objects \citep{bell11}, or meet limited success in RQ ones \citep{jones17}.
No such campaign, as far as we know, exists for higher frequencies. \\

In this paper, we report results from a dedicated monitoring campaign of \ngc\ 
with the single 30\,m dish of the Institut de Radioastronomie Millimetrique (\iram ) at 95 GHz and at 143\,GHz, simultaneous with \xrt\ and \uvot\ monitoring that was part of a multi-wavelength campaign during the last part of 2015 \citep{Behar17}.
The \iram\ observations were subsequently obtained to match, as closely as possible, the dense part of the \xrt\ monitoring period. 

\section{Observations}
\label{sec:data}

\subsection{\ngc }
\ngc\ is a nearby Seyfert galaxy at $z=0.01588, D_L=71.2$\,Mpc, where $1\arcsec$ corresponds to $\approx300$\,pc. 
It was observed many times with the VLA from 1.4\,GHz up to 22\,GHz \citep{unger87, Lal04, Orienti10, Prouton04}, with the \vlbi\ \citep{Lonsdale93, Lonsdale03} and with \merlin\ \citep{Thean01}.
At 95\,GHz it was observed with \carma\ \citep{Behar15, baldi15}, and at 350\,GHz with \alma\ \citep{izumi15}.

The high X-ray brightness and variability of \ngc\ make it a popular target for X-ray telescopes, and in particular monitoring \citep[e.g.,][]{markowitz04}.
In the visible too, \ngc\ was monitored extensively, notable examples of which are its wavelength-dependent continuum delays of several light days, which indicate processing in an accretion disk \citep{Collier98}, and reverberation mapping resulting in a black-hole mass measurement of $10^7M_\odot$  \citep{Peterson14}.
In the following we describe the observations of our 2015 campaign.

\subsection{mm-wave}
\ngc\ was monitored daily with the EMIR multi-band mm-wave receiver on the \iram\ 30\,m telescope from 2015, Nov 15 through 2015, Dec 31, except for 1 week at the end of November when an EMIR upgrade took place.
Each observation included about 1\,hr on source, which provides a sensitivity of typically 0.15\,mK (0.8\,mJy 1$\sigma$).
In total, we obtained 36 flux measurements.
The campaign yielded light curves in four EMIR bands \citep[within B1 and B2, see][]{EMIR12}, each 8\,GHz wide.
These are listed in Table\,\ref{tab:EMIR}. 

\begin{table}
\centering
\caption{\iram /EMIR bands used in this work.}
\begin{tabular}{lcc} 
Name  & Lower bound & Upper bound \\ 
  &  (GHz)  &  (GHz) \\ 
 \hline
E090L & 83 & 91 \\
E090U & 99 & 107 \\
E150L & 131 & 139 \\
E150U & 147 & 155 \\
E090 & \multicolumn{2}{c}{E090L \& E090U} \\
E150 & \multicolumn{2}{c}{E150L \& E150U} \\
\hline
\end{tabular}
\label{tab:EMIR}
\end{table}

Antenna temperatures were obtained after subtracting the sky background measured by wobbling the secondary mirror at 2\,Hz.
To separate the effect of the optics between the two wobbler positions, we perform sub-scans with the source in one position of the wobbler,
and then in the other position, and average the two. 
We then convert antenna temperature, after correction for atmospheric attenuation, to flux density using 6.2Jy/K at 86\,GHz, 8.8Jy/K at 230\,GHz, and a linear interpolation in between. 
We used Uranus as the flux calibrator. It was low on the horizon, and observed at 8.3\,Jy. 

\begin{figure}
\vskip -3.cm
\includegraphics[scale=0.6,angle=0]{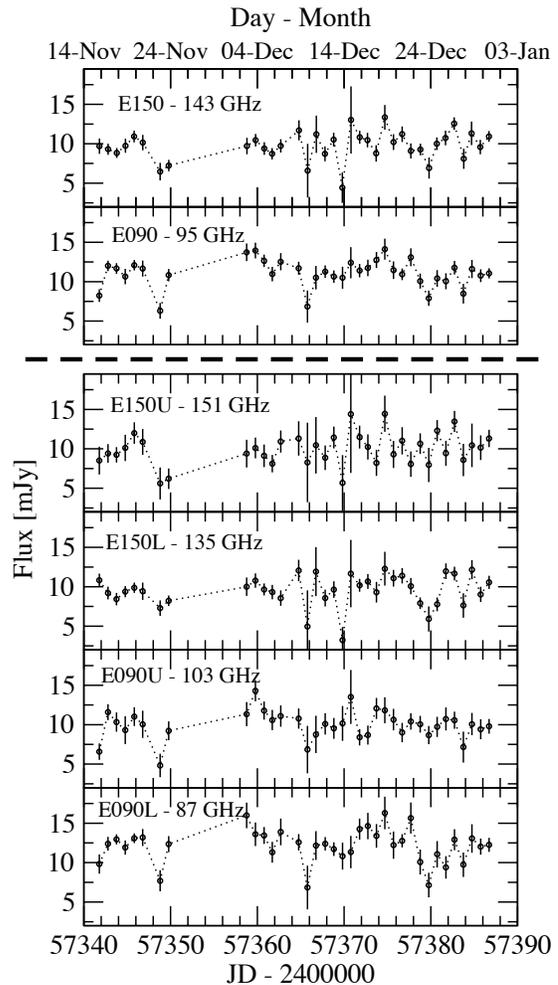}
\vskip -0.3cm
\caption{\iram\ light curves in the four EMIR bands; central frequency is indicated for each (see Table\,\ref{tab:EMIR} for details).
The top two are co-added pairs from the bottom four.
All panels span the same vertical range. Lines are drawn just to guide the eye. 
}
\vskip -0.35cm
\label{fig:IRAM_lc}
\end{figure}

In each band, we averaged over the two polarizations. 
Fig.~\ref{fig:IRAM_lc} shows the four light curves in the four bottom panels.
Although the photometric uncertainties are large  ($1\sigma = 0.8$\,mJy), in some cases there is inter-day variability of more than 1$\sigma$.
We then also averaged each two adjacent bands to improve the Signal to Noise Ratio (S/N).
The two averaged bands are (somewhat confusingly) named E090 centered around 95\,GHz, and E150 centered around 143\,GHz.
Their light curves are shown in the two top panels with their smaller error bars, where inter-day variability can exceed 2$\sigma$.

The flux density of \ngc\ at 95\,GHz and at 143\,GHz ranges between 6--14\,mJy, and 4--13\,mJy, respectively, which corresponds to 
$\nu L_\nu =$ $(3.5- 8.1)\times 10^{39}$\,erg\,s$^{-1}$ at 95\,GHz, and $\nu L_\nu =$ $(3.5- 11.3)\times 10^{39}$\,erg\,s$^{-1}$ at 143\,GHz.
The present 95\,GHz flux densities are marginally consistent with previous \carma\ measurements, where the beam was much smaller.
In the C configuration we measured 5.0\,mJy in Nov 2013 \citep{Behar15}, and 2--4\,mJy (6--7\,mJy in the D configuration) in Mar-Apr 2014 \citep{baldi15}.
See Fig.\,\ref{fig:SED} for a graphical comparison.
The total- (as opposed to peak-) flux densities in 2014 are 6--11\,mJy in the C configuration and 12--15\,mJy in the D configuration \citep{baldi15}.
The 30\,m telescope has a beam size of $28 \arcsec$ at 90\,GHz and $17 \arcsec$ at 150\,GHz, while \carma\ had a beam size at 95\,GHz of 2\farcs2 in the C configuration and 5\farcs5 in the D configuration.
The 1\,kpc circumnuclear starburst ring of \ngc\ is 2\farcs9 across \citep{Diaz-Santos07}.
The reasonably good agreement between \iram\ and \carma , given also the variability, indicates the majority of the mm-wave flux comes from the AGN, and not from stars.
Much below these length scales, however, any extended source beyond 1\,light-month would dilute the variability signal from the core that we are after.

\subsection{Radio-mm SED}
\begin{figure}
\vskip -0.4cm
\hskip -1.cm
\includegraphics[scale=0.41,angle=0]{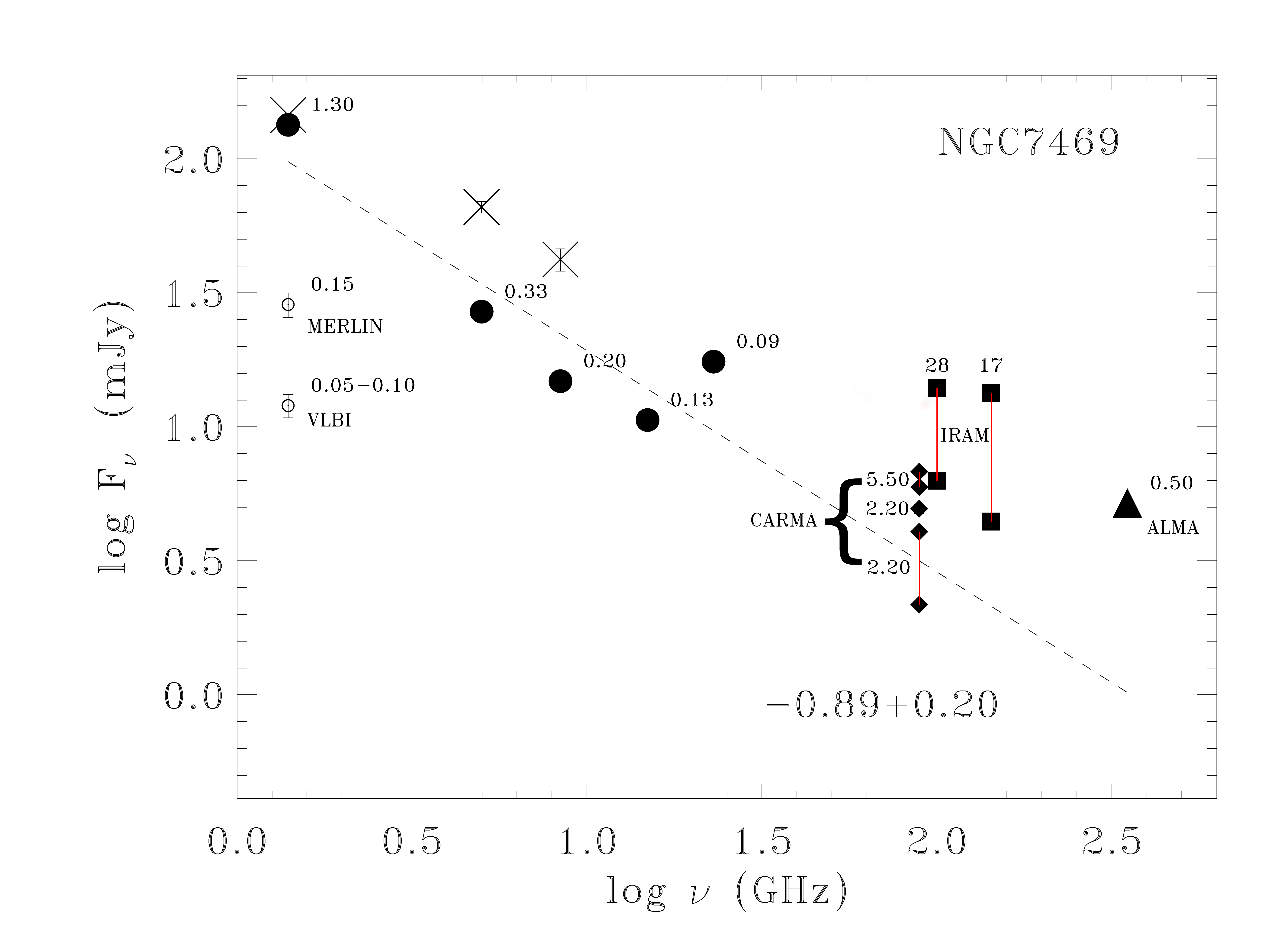}
\vskip -0.4cm
\caption{Radio to mm SED of \ngc . Numbers indicate beam size in arcseconds. 
VLA high-resolution data (A array) are plotted as full circles, lower resolution data (B-C-D arrays) as {\large $\times$}'s.
Very high resolution data (\merlin\ and \vlbi ) are shown with hollow circles.
In the mm, present \iram\ data are represented by squares, earlier \carma\ data by diamonds; red lines indicate the range of variability in each campaign (see text). 
The \alma\ measurement is the triangle. 
Fitting the A-array VLA data results in the dashed-line spectral slope of --0.89.
\iram , \alma , and some \carma\ points at $\gtrsim 90$\,GHz fall above this slope, and may therefore represent a different AGN spectral component. 
}
\vskip -0.2cm
\label{fig:SED}
\end{figure}

In Fig.\,\ref{fig:SED} we plot the full radio-mm spectral energy distribution (SED) of \ngc , based mostly on archival measurements.
Previous VLA observations were detailed in \citet{Behar15}.
The present 95\,GHz and 143\,GHz data points (squares) can be seen to lie above the low-frequency steep slope of $\alpha = -0.89$.
They are also much above the extrapolation from 350\,GHz of the steep thermal dust emission in the far infrared (FIR) \citep{izumi15}.

At low radio frequencies, \ngc\ is very bright, e.g., $F_{\rm 1.4GHz} =134$\,mJy \citep{unger87}, which poses a concern of stellar contamination. 
The \vlbi\ map of the inner region at 1.67\,GHz shows several resolved and unresolved components associated with the AGN, each of order 10\,mas (i.e., 3\,pc) in size; the identification of the central AGN, however, is unclear \citep[][Fig.\,2 therein]{Lonsdale03}.
Each of these point sources emits a few mJy for a total of $\sim$15\,mJy in the central $\sim 0\farcs1$ \citep[][]{Lonsdale93,Lonsdale03}. 
\vlbi\ apparently is insensitive to additional diffuse 15--20\,mJy in that field, which \merlin\ does detect within $\sim0\farcs15$ \citep{Thean01,Lonsdale03}.
Interestingly, these are also the scales of the strong X-ray outflow of \ngc , which is estimated to be around 30\,pc or less \citep{Peretz18, Mehdipour18}. Thus, extended, optically-thin synchrotron emission is in fact expected from that region.
If \ngc\ varies at 95\,GHz on time scales of days to weeks, as presently monitored, this mm component would be distinct from the low-frequency component, which seems to be the case, based on the SED (Fig.\,\ref{fig:SED}) as well. 

At higher frequencies, $F_{\rm 350GHz}= 5.19\pm0.09$\,mJy was measured with ALMA from an unresolved $0\farcs5$ core \citep{izumi15}, which those authors conclude is due to thermal dust, because it is much above the extrapolated steep-slope power law. 
However, the 350\,GHz flux may also contain a contribution from the optically thick mm-wave spectral component that is dominant at 100\,GHz.
Additionally, \citet{izumi15} measure an internal slope around 350\,GHz of $\alpha = +2\pm1$, while dust would usually have $\alpha = 3.5$.
Extrapolating this slope down to the present band of 90 -- 150\,GHz yields flux densities that are much below the measured 10 -- 11\,mJy, even for $\alpha = 1$.

A similar SED, with mm-wave excess above the low-frequency steep slope, and above the modeled dust from the FIR, is observed for another Seyfert galaxy {\small NGC\,3227} \citep[][Fig.\,4 therein]{Alonso-Herrero19}, whose ALMA observations show a rather complex structure of different spectral slopes $-1 < \alpha < +2$ between 230 -- 350\,GHz in the central few arcseconds (Fig.\,3 therein).
This likely indicates several components are contributing to the core emission, both in  {\small NGC\,3227} and in \ngc .

\subsection{\it Swift}

During the entire mm-wave monitoring period,
 \ngc\ was also observed daily for 1500\,s with the X-ray Telescope ({\small XRT}) 
 and the UV/Optical Telescope ({\small UVOT}) on board \swift .
Exposures with each of the six {\small UVOT} filters are for 5\,min.
Data were reduced using xselect within the heasoft software package\footnote{https://heasarc.gsfc.nasa.gov/lheasoft/}.
{\small XRT} was operated in PC mode.
Source counts were extracted from a 20\,pixel radius circle around the source and background from an annulus up to 30 pixels.
{\small XRT} uncertainties are dominated by count statistics and are typically only a few percent.
The light curves from all \swift\ instruments are presented in Fig.~\ref{Swift_lcs}. 
The X-rays, measured between 0.3-10\,keV can be seen to vary dramatically with no obvious trend, while the six UV/Optical light curves feature a rather uniform, slow and monotonous increase with time over the monitoring period.

\subsection{\it \xmm\ Optical Monitor}

\begin{figure}
\includegraphics[scale=0.33,angle=0]{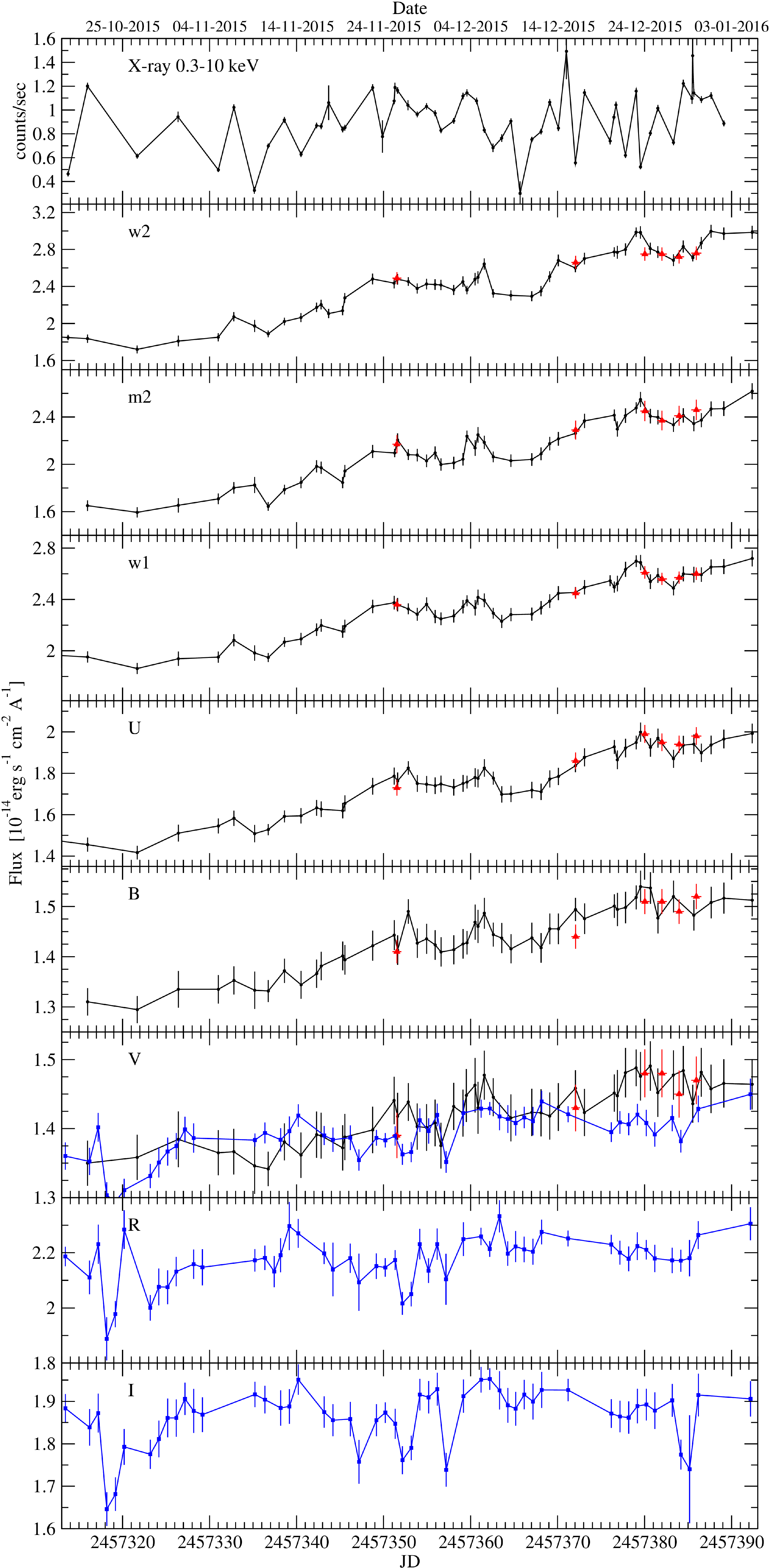}
\caption{\swift\ light curves in X-rays and in six UV-optical bands (black points). 
Red triangles indicate (6) epochs measured with the \xmm\ OM.
Monitoring from the ground with the Wise Observatory is shown in the bottom three panels (blue squares).
Lines are drawn just to guide the eye.
All the data are listed in the appendix.}
\label{Swift_lcs}
\end{figure}

\xmm\ observed \ngc\ six times during the \iram\ monitoring period.
\ngc\ varies in X-rays within hours, hence \xmm\ probes much shorter time scales than we can currently monitor in the mm with \iram .
The \xmm\ X-ray spectra were described in detail elsewhere \citep{Behar17, Peretz18}.
Here, we focus on the \xmm\ {\small Optical Monitor} (OM).
The OM observed \ngc\ for 4\,ks with each of its six UV and optical filters, which are the same as those of \uvot . 
OM data were reduced using the XMM Science Analysis System (SAS) pipeline
and the measurements are presented in Fig.\,\ref{Swift_lcs} as red points. 
As expected, there is very good agreement between \uvot\ and \xmm /OM in all filters.

\subsection{\it Wise Observatory}

Optical observations in Bessel V, R, and I filters were carried out with the 28$^{\prime\prime}$
Jay Baum Rich telescope at the Wise Observatory, Israel. 
An FLI ProLine PL16801 CCD was used as a detector with a 
pixel scale of 0.834 arcsec/pix. \ngc\ was observed for 5 min in each filter once a night.
On some nights several exposures were carried out in each filter and averaged into one data point per night. 
The data were reduced using standard IRAF\footnote{IRAF
(Image Reduction and Analysis Facility) is distributed by the National
Optical Astronomy Observatories, which are operated by AURA, Inc,
under cooperative agreement with the National Science Foundation.}
procedures. The broad-band light curves in the three filters
are produced by comparing their instrumental magnitudes to those of
constant-flux stars in the field \citep[e.g.,][]{Netzer96}.
Absolute photometry calibration was achieved by using stars with known magnitudes in the field of \ngc . 
The Wise light curves are shown in the three bottom panels of Fig.\,\ref{Swift_lcs} with blue data points; 
the V band is over-plotted on the \swift\ OM V band, for comparison. 
We suspect these ground observations, when (slightly) discrepant from the space observations, are flawed due to atmospheric effects.

\section{Results and Discussion}
\label{sec:results}

\subsection{Variability}
\label{sec:variability}

The summary of the variability parameters for the various light curves is given in Table~\ref{tab:variability}.
For each light curve we list the mean flux density $\langle F_\nu \rangle$ and its standard deviation $STD$.
The $\chi ^2$ of the light curve with respect to a constant $\langle F_\nu \rangle$ is also listed.
In the \iram\ light curves there are $N=36$ data points, hence $N-1= 35$ degrees of freedom for the reduced $\chi ^2$.
The associated p-values, computed from $\chi ^2$ and the degrees of freedom, give an idea of the significance of the variability in each light curve, 
and can be thought of as the probability for no-variability, given the data and $\chi ^2$.
These p-values in the \iram\ light curves range from $10^{-7}$ to 0.1.
The clearest indication of variability comes from the combined 95\,GHz light curve E090.
The least significant variability is in the E150U (151\,GHz) light curve, but after combining with E150L, to give E150 (143\,GHz), 
it indicates variability is highly likely ($p= 3\times 10^{-5}$).
Computing the various mm band correlations among themselves, yields nothing but a peak at zero, 
indicating this entire band, which we refer to loosely as the AGN mm-wave component, more or less varies together.
The significance of the present variability is comparable and even higher (with twice as many data points) than that of the peak-flux light curve measured with \carma\ \citep[][see also Table\,\ref{tab:variability} below]{baldi15}, despite the order of magnitude disparity in beam size, which implies the dominant variability of the unresolved core.


\begin{table*}
 \centering
\caption{Variability parameters of the various light curves of \ngc .}
\begin{tabular}{lcccccccc}
\hline
\hline
Light curve  &  $N$ & $\nu $ & $\langle F_\nu \rangle$ &  $STD$ &   $\fv $ & $\frac{F_\nu^\mathrm{max}}{F_\nu^\mathrm{min}}$ & $\chi ^2$ & p-value\\
 & &  & & &  (\%) & & & \\ 

\hline
\iram\ (Fig.\,\ref{fig:IRAM_lc})  & & (GHz) & (mJy) & (mJy) &  & & & \\
E150	& 36	& 143 & 9.7	& 1.9	& 12.6$\pm$3.4	& 3.0	& 78.6	& 3$\times10^{-5}$ \\
E090	& 36	& 95	& 11.1	& 1.8	& 12.9$\pm$2.4	& 2.2	& 96.9	& $10^{-7}$	\\
E150U	& 36	& 151 & 9.9	& 2.1	& 0	$^1$	        & 2.6	 	& 45.8	& 0.1	\\
E150L	& 36	& 135 & 9.5	& 2.1	& 14.0$\pm$3.9	& 3.8	& 81.6	& $10^{-5}$	\\
E090U	& 36	& 103 & 10.0	& 1.8	& 9.7$\pm$4.1		& 3.0	& 56.5	& 0.01	\\
E090L	& 36	& 87	 & 12.1	& 2.2	& 13.5$\pm$2.9	& 2.4	& 66.7	& $10^{-3}$ \\
\hline
\multicolumn{2}{l}{\carma\ \citep{baldi15} } \\
$F_\nu^\mathrm{peak}$  & 18 &  95 &   3.0 & 0.5  & 12.8$\pm$3.6   &  1.9  &  45.5 & 2$\times10^{-4}$ \\
$F_\nu^\mathrm{tot}$   & 18 &  95 &   8.1 & 1.0  & 7.4$\pm$3.5  & 1.6 &   22.4 & 0.17 \\
\hline
\xrt\ (Fig.\,\ref{Swift_lcs}) &  & $h\nu$\,(keV) & (counts/s) & (counts/s) &   & & & \\
Total & 57  & 0.3-10.0 & 0.91 & 0.23 & 25.0$\pm$2.6  & 5.0 & 3172.8 & 0\\
Hard & 57	& 2.0-10.0	& 0.37	& 1.00	& 24.3$\pm$2.6 & 4.4 & 1746.9 & 0\\
Soft	& 57	& 0.3-2.0	& 0.54	& 0.16	& 28.4$\pm$2.9 & 5.7 & 2768.9 & 0 \\
Hardness Ratio	& 57	& \nodata &$\left<HR\right>=-0.17$	& 0.06	& 0 $^1$	& \nodata 	& 303.4& 	$10^{-35}$\\
\hline
\uvot\ (Fig.\,\ref{Swift_lcs})	 &  & $\lambda$\,(\AA ) & (10$^{-14}$erg\,s$^{-1}$cm$^{-2}$\AA$^{-1}$) & (10$^{-14}$erg\,s$^{-1}$cm$^{-2}$\AA$^{-1}$) &   & & & \\
w2	& 54	& 1928	& 2.44	& 0.35	& 14$\pm$1.4	& 1.7	& 2651.9	& 0	\\
m2	& 49	& 2246	& 2.12	& 0.26	& 12$\pm$1.3	& 1.6	& 1249.7	& 0	\\
w1	& 53	& 2600	& 2.34	& 0.22	& 9.1$\pm$0.9	& 1.5	& 1045.0	& 0	\\
U	& 52	& 3465	& 1.76	& 0.15	& 8.1$\pm$0.9	& 1.4	& 792.5		& 0\\
B	& 48	& 4392 	& 1.43	& 0.06	& 3.9$\pm$0.5	& 1.2	& 235.7	& 6$\times 10^{-27}$\\
V	& 49	& 5468	& 1.42	& 0.04	& 1.8$\pm$0.5	& 1.1	& 80.7	& 2$\times 10^{-3}$	\\
\hline
Wise	 (Fig.\,\ref{Swift_lcs})	 
\\								
V	& 51	& 5510	& 1.39	& 0.04	& 2.1$\pm$0.3	& 1.1	& 193.1 & 	$10^{-18}$ \\
R	& 52	& 6580	& 2.18	& 0.09	& 3$\pm$0.5	& 1.2	& 133.4	 & 3$\times 10^{-9}$ \\
I	& 52	& 8060	& 1.86	& 0.07	& 2.8$\pm$0.5	& 1.2	& 159.3 & 	4$\times 10^{-13}$ \\
\hline
\hline
\end{tabular}
\raggedright
\tablenotetext{1}{See text after Eq.\,\ref{eq:fvar}.}
\tablecomments{The data of the light curves are listed in their entirety in the appendix.}
\label{tab:variability}
\end{table*}

The amplitude of variability in a light curve with $N$ data points can be quantified with $\fv$ \citep[e.g.][]{edelson02,vaughan03}:

\begin{equation}
\fv = \sqrt{STD^2 - \langle \sigma ^2 \rangle \over \langle F_\nu \rangle^2},
\label{eq:fvar}
\end{equation}

\noindent and its uncertainty $\sigma_{\fv}$

\begin{equation}
\sigma_{\fv} = \sqrt{  \left\{ \sqrt \frac{ \langle\sigma ^2\rangle}{N} \cdot \frac{1}{\langle F_\nu \rangle} \right\} ^2         +       \left\{ \sqrt \frac{1}{2N} \cdot \frac {  \langle\sigma ^2\rangle }{\langle F_\nu \rangle^2 \fv} \right\} ^2             }
\end{equation}

\noindent $\fv$ represents the intrinsic variability amplitude in excess of the mean measurement uncertainty $\left< \sigma \right>$.
When the latter exceeds the former, $\fv$ is not well defined, and a value of zero is given in Table\,\ref{tab:variability}.
The $\fv $ values of the combined light curves E090 and E150 over our 45 day  \iram\ campaign are both at the $13\% \pm 3\% $ level, very close to the previous \carma\ results. 
$\fv $ of the 0.3-10\,keV X-ray light curve of \xrt\ is twice as high at $25\% \pm 3\% $, but covers 75 days.
For a better comparison with \iram\ we therefore computed $\fv$ for all \xrt\ subsets of 45 days, and find $\fv = 23\% - 26\%$ for all, so no difference from 75 days.

\citet{markowitz04} measured $\fv$(2-12\,keV) to be 16.0\%$\pm$0.4\% for \ngc\ over 36 days using light curves from the {\small RXTE} telescope.
We computed $\fv$ for all present subsets of 36 days, which have at least 20 data points, and still get $\fv = 22\% - 26\%$.
Apparently in our \xrt\ light curve, $\fv$ is the same for any time scale between 30 - 70 days.
This is consistent with \citet[][their Fig.\,4]{Vagnetti16}, who found excess variance to vary slowly with light curve duration, as $t^{0.2}$.
The higher $\fv$ values in the present work than in \citet{markowitz04} are likely due to the more precise (lower $\left< \sigma ^2 \right>$) measurements of \xrt , and to the stronger variability in the soft X-ray band. See Sec.\,\ref{sec:X-ray} for more discussion on the hardness ratio. 

Yet another measure of variability amplitude is the ratio of the highest $F_\nu^\mathrm{max}$ to lowest $F_\nu^\mathrm{min}$ flux density (Table\,\ref{tab:variability}).
In the present \iram\ data this ratio is between 2 and 4, while those of \carma\ are less than 2 (Table\,\ref{tab:variability}). 
This may not be expected given the larger \iram\ beam size, if there was appreciable extended emission, but evidently much of the emission is within a few arcseconds.
The X-ray variability amplitude is even higher, reaching a factor of 5, while it is less than a factor 2 in the UV and down to 10\% in the visible band.
Interestingly, strong mm flux variability by a factor of 2 is observed within a day (Fig.\,\ref{fig:IRAM_lc}).
This suggests a 95\,GHz source size of no more than 1 light day, which is just less than $10^{-3}$\,pc, and which is approximately the physical size of the \ngc\ core estimated from the luminosity of a self absorbed synchrotron source \citep{Behar15}.
These results call for intra-day mm-wave monitoring, and even better, coordinated mm-wave and X-ray monitoring on these shorter time scales.

Variability in the IR through the UV was measured in Seyferts many times, and specifically in \ngc .
This variability for the most part is of small amplitude, as in Fig.~\ref{Swift_lcs}, 
clearly distinct from that of the X-rays and the mm bands, which therefore likely arise from a different AGN component.
Nonetheless, monitoring over many years reveals occasional significant drops in flux within months.
Interestingly, in the IR the longest wavelengths (L band) vary most, and in the optical it is the shortest wavelengths (U band) \citep{Glass98}. 
This is likely because those arise from and gas closer in to the center, but the time scales imply it is still much farther away than the X-ray and mm sources.


 
\subsection{X-ray Variability}
\label{sec:X-ray}

\begin{figure}
\includegraphics[scale=0.35,angle=0]{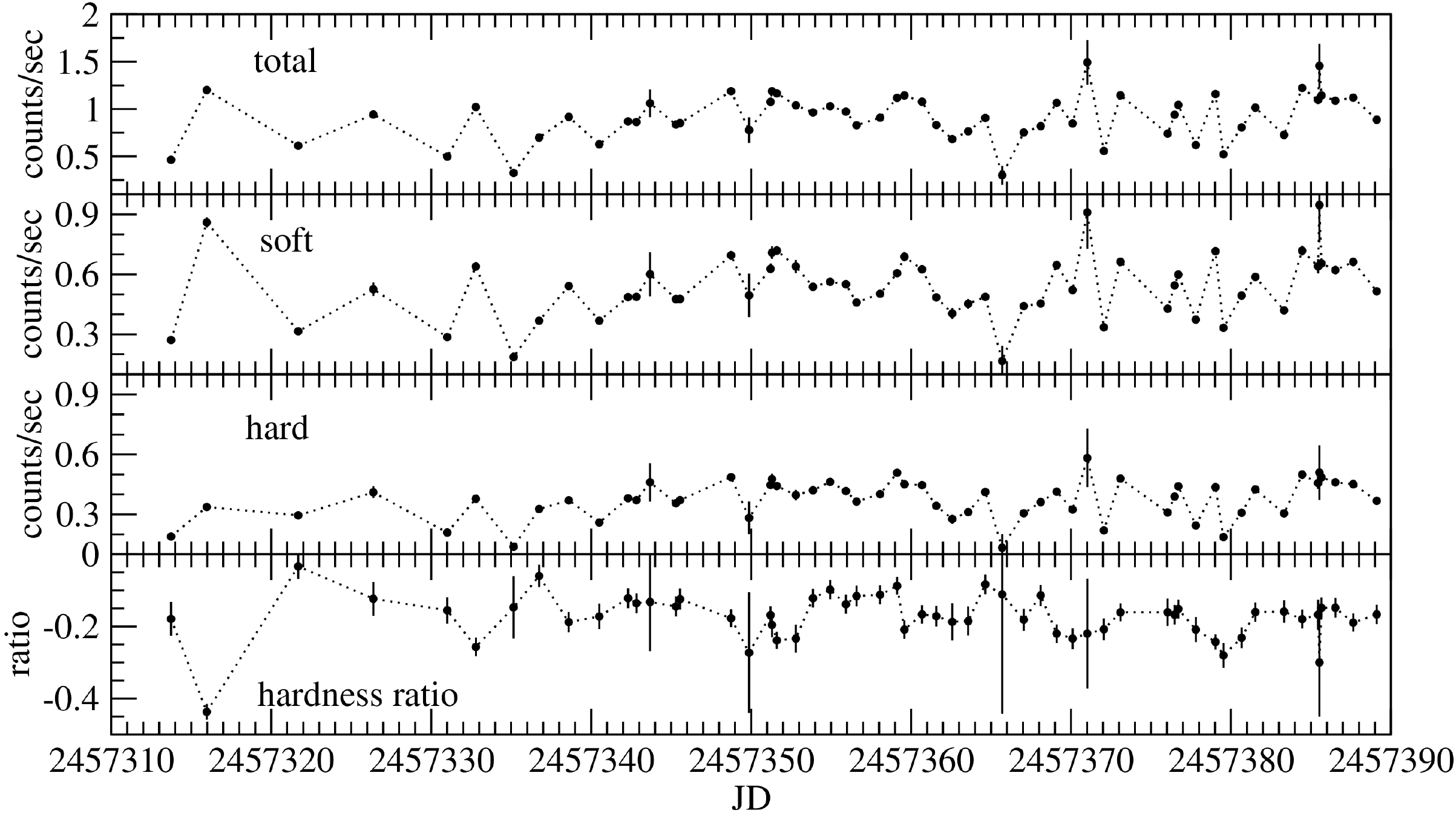}
\caption{X-ray light curves of  \ngc\ in the hard (2.0-10.0\,keV), soft (0.3-2.0\,keV), and full bands. 
The light curve of the hardness ratio $HR=(H-S)/(H+S)$ is also plotted.}
\label{fig:X-ray}
\end{figure}

\xrt\ covers the broad energy band from 0.3 to 10.0~keV.
It is beneficial to study the  hard (2.0-10.0\,keV) and soft (0.3-2.0\,keV) bands separately, as they could come from separate spectral components of the corona.
The hardness of the spectrum can be an indication of the physical state of the source, 
and is therefore often also measured for its variability \citep{McHardy99, McHardy06, Peretz18b}.
We denote here the hardness ratio in terms of \xrt\ count rates (counts/s) $HR = (H-S)/(H+S)$, where $H$ and $S$ are the count rates in the hard and soft bands, respectively.
Consequently, $-1 < HR < 1$.
In Fig.~\ref{fig:X-ray}, we present the \xrt\ light curves in the two separate bands as well as that for $HR$.
From inspection of Fig.\,\ref{fig:X-ray} and Table\,\ref{tab:variability} it can be seen that the soft band is slightly more variable.

X-ray variability can be caused by varying photo-electric absorption, rather than by genuine coronal variability.
In that case, the source would become harder when dimmer \citep{Mehdipour17}. 
This behaviour is not seen in the present case, where both bands vary relatively in unison, and $HR$ varies relatively little.

Recently, \citet{Mehdipour19} found in a sample of RL AGNs that radio loudness scales inversely with X-ray column density, so those sources are indeed more absorbed in X-rays (harder) when dimmer in the radio.
Following that work, we checked for a correlation between the present E090 light curve (Fig.\,\ref{fig:IRAM_lc}) and the X-ray hardness-ratio light curve (Fig.\,\ref{fig:X-ray}), which could be a proxy of absorption.
We find only a weak correlation (Cross Correlation Coefficient $CCF\gtrsim0.5$) at $t=0$.
If significant, this would indicate the mm-flux increases at the same time the X-rays become harder, which is opposite the effect of RL AGN \citep{Mehdipour19}.
In the context of a magnetic AGN corona, the correlation can be explained by freshly generated hot electrons.
These electrons increase the synchrotron (mm) on one hand, and the most Compton boosted X-rays (hardness) on the other hand.
However, such correlations and inferences need to await better short-temporal sampling of the mm and X-ray light curves.

\subsection{The mm X-ray Connection}
\label{sec:correlation}

\begin{figure}
\includegraphics[scale=0.33,angle=0]{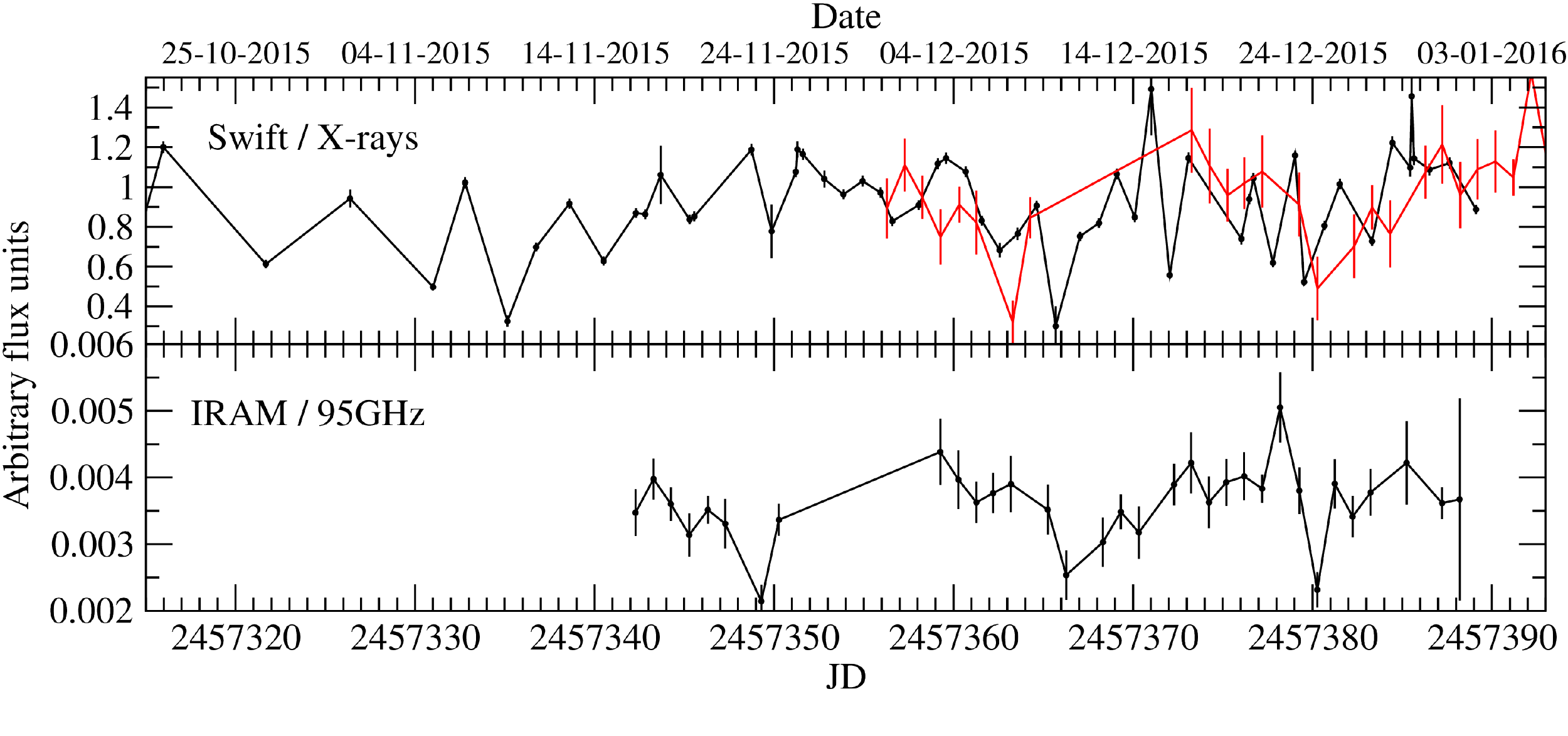}
\caption{Light curves of \ngc\ in X-rays (top) and 95\,GHz (bottom). 
A tentative 14 day lag is demonstrated in the top panel by the red points, which are the (scaled) 95\,GHz light curve shifted forward by 14 days.}
\label{fig:combined}
\end{figure}

Among the \iram\ light curves, E090 (95\,GHz) shows the most robust variability.
Hence, the E090 and X-ray light curves are compared in Fig.~\ref{fig:combined}.
The light curves are different, though the variability time scale of one-to-few days appears to be similar.
In order to check for any time-lag between the two, 
we try to cross correlate the X-ray light curve with the mm one. 
No clear correlation is found, except a weak peak  of $CCF=0.6$ at $-14$ days. 
In the top panel of Fig.\,\ref{fig:combined} we plot in red the 95\,GHz light curve shifted forward by 14 days, on top of the X-ray one.
It shows the gradual decrease in X-ray flux during the third week of December, followed by a gradual increase in the fourth week.
The mm-wave light curve gives the sense of going through a similar transition two weeks earlier.
Shorter time scales do not seem to correlate, but simultaneous monitoring on time scales of hours is required before any conclusive statement can be made.

The notion of radio/mm flares preceding X-ray flares can be associated with acceleration of fast electrons that emit synchrotron in the vicinity of magnetic fields, subsequently depositing their energy in the ambient medium, which then emits an X-ray flare.
This is the scenario in some stellar coronal flares:
Electrons accelerated through magnetic reconnection in the corona emit synchrotron radio, but X-rays are emitted only after they thermalize as they reach and evaporate the much denser chromosphere. 
In this scenario, only a small fraction of the energy is emitted in the radio/mm band.
The vast majority is emitted eventually in X-rays, hence $L_{\rm X} \gg L_{\rm R}$.

Fig.\,\ref{fig:combined} is only suggestive of such a connection in \ngc , but a time lag of a few days is interesting.
In stellar coronae the lag between radio and X-rays can be $\sim$hours, which is by far longer than the coronal crossing time of $\sim$seconds, and not well understood.
This means the electrons are trapped in the entangled coronal magnetic fields and diffuse slowly down to the chromosphere \citep[for more details, see][]{guedel02}. 
Following through with the AGN analogy, if in \ngc\ the lag is a few days and up to two weeks, the lifetime of fast electrons in the magnetic structures in the accretion disk corona would be about two orders of magnitude longer than they are in stellar coronae.
If this is due to the size of the magnetic structures, scaling from stellar radii yields structures of $10^{13}$\,cm, or in \ngc\ a few gravitational radii, which is presumably the size of the X-ray corona.
An important difference between stellar and AGN corona is the X-ray emitting processes.
While in stellar corona the X-rays are emitted by thermal ($10^6$K) ions, in AGN X-rays are generated by ($10^9$K) electrons through Compton up-scattering of accretion-disk photons.

\section{Summary and Conclusions}
\label{sec:summary}
 
We monitored \ngc\ at 95\,GHz and at 143\,GHz with the \iram\ 30\,m telescope over approximately a month and a half.
The measured light curves indicate variability on time scales of days and less, similar to that detected before in \ngc\ with \carma\ \citep{baldi15}.
The amplitude of variability is also similar in both campaigns, despite the much larger beam sizes in the present work, indicating the source is dominated by a small-size (light days) core.

In the present work, for the first time we measured an X-ray light curve simultaneous with the mm-wave ones.
Both mm and X-ray sources vary on a similar time scale of days. 
This is in sharp contrast with the simultaneous UV-Optical light curves that slowly increased in flux in eight different filters during the course of the campaign, with no evident daily variability.
This temporal behaviour constitutes clear evidence that the mm-wave source can be associated with the X-ray source (i.e., corona), but is distinct from the UV-Optical source (the accretion disk).
The amplitude of variability over measurement uncertainties and its significance is somewhat higher in X-rays than in mm-waves, mostly due to the much more precise photometric measurements in X-rays.
A tentative correlation between the mm and X-ray light curves suggests a $\sim 14$ day X-ray lag.
If real, this lag could imply, by analogy to stellar coronae, that magnetically heated electrons produce the mm waves, and ultimately lose most of their energy by emitting X-rays.

Clearly, the present evidence for a temporal correlation, and for a true mm-wave flare preceding an X-ray one is only suggestive.
Such a tentative connection between the mm-wave and X-ray sources, however, begs for more monitoring of nearby AGN, simultaneously in both bands.
Since both the mm and the X-ray sources seem to vary on time scales which could be shorter than 1 day, it would be beneficial to monitor on both inter- and intra- day time scales.
Such campaigns require higher sensitivity than that available with the \iram\ 30\,m telescope.
Thus, higher-sensitivity (and resolution) arrays, e.g. \alma , \iram{\small /NOEMA}, and later SKA would be more suitable. 

\section*{Acknowledgments}
This research is supported by the I-CORE program of the Planning and Budgeting Committee (grant 1937/12),
and was performed in part at Aspen Center for Physics, which is supported by National Science Foundation grant PHY-1607611. 
We thank an anonymous referee for useful comments.

\bibliography{my}

\appendix
\section{Light curve data}
The complete data sets of the present monitoring campaign are listed below.

\begin{landscape}
 \begin{table}
  \caption{\swift\ light curves. 
  {\small XRT} measurements are in counts/s.
  All {\small UVOT} flux densities $F_\lambda$ are in $10^{-14}$erg\,s$^{-1}$\,cm$^{-2}$\AA$^{-1}$. 
  JD is listed as JD-2457300.}
  \label{tab:lightcurves1}
\scriptsize
  \begin{tabular}{cccccccccccccc}
    \hline
    \multicolumn{14}{c}{\swift } \\
    \multicolumn{2}{c}{\small XRT} & \multicolumn{2}{c}{w2} & \multicolumn{2}{c}{m2} & \multicolumn{2}{c}{w1} & \multicolumn{2}{c}{U} & \multicolumn{2}{c}{B} & \multicolumn{2}{c}{V}  \\
    JD & counts/s & JD & $F_\lambda$ & JD & $F_\lambda$ & JD & $F_\lambda$ & JD & $F_\lambda$ & JD & $F_\lambda$ & JD & $F_\lambda$   \\
\hline
13.740 & 0.464$\pm$0.022 & 13.741 & 1.848$\pm$0.029 & 15.974 & 1.651$\pm$0.045 & 15.976 & 1.952$\pm$0.045 & 15.979 & 1.455$\pm$0.034 & 15.982 & 1.310$\pm$0.027 & 15.971 & 1.350$\pm$0.033 \\ 
15.975 & 1.201$\pm$0.029 & 15.968 & 1.836$\pm$0.045 & 21.686 & 1.594$\pm$0.043 & 21.689 & 1.862$\pm$0.043 & 21.692 & 1.417$\pm$0.033 & 21.695 & 1.295$\pm$0.027 & 21.683 & 1.358$\pm$0.033 \\ 
21.687 & 0.613$\pm$0.021 & 21.681 & 1.720$\pm$0.042 & 26.413 & 1.653$\pm$0.060 & 26.414 & 1.938$\pm$0.056 & 26.415 & 1.511$\pm$0.041 & 26.416 & 1.335$\pm$0.037 & 26.412 & 1.384$\pm$0.040 \\ 
26.382 & 0.942$\pm$0.044 & 26.411 & 1.810$\pm$0.058 & 31.003 & 1.708$\pm$0.045 & 31.006 & 1.950$\pm$0.044 & 31.009 & 1.546$\pm$0.036 & 31.012 & 1.335$\pm$0.028 & 31.000 & 1.365$\pm$0.033 \\ 
31.004 & 0.498$\pm$0.018 & 30.997 & 1.851$\pm$0.045 & 32.793 & 1.801$\pm$0.048 & 32.796 & 2.083$\pm$0.047 & 32.799 & 1.583$\pm$0.037 & 32.802 & 1.353$\pm$0.028 & 32.790 & 1.367$\pm$0.033 \\ 
32.794 & 1.022$\pm$0.027 & 32.787 & 2.072$\pm$0.050 & 35.181 & 1.825$\pm$0.068 & 35.182 & 1.983$\pm$0.059 & 35.182 & 1.508$\pm$0.042 & 35.183 & 1.333$\pm$0.037 & 35.180 & 1.346$\pm$0.041 \\ 
35.152 & 0.325$\pm$0.029 & 35.179 & 1.972$\pm$0.065 & 36.744 & 1.644$\pm$0.037 & 36.746 & 1.948$\pm$0.035 & 36.747 & 1.528$\pm$0.027 & 36.749 & 1.332$\pm$0.022 & 36.742 & 1.342$\pm$0.025 \\ 
36.745 & 0.698$\pm$0.021 & 36.741 & 1.887$\pm$0.037 & 38.604 & 1.786$\pm$0.041 & 38.606 & 2.068$\pm$0.039 & 38.607 & 1.592$\pm$0.029 & 38.608 & 1.372$\pm$0.024 & 38.603 & 1.381$\pm$0.027 \\ 
38.603 & 0.917$\pm$0.026 & 38.601 & 2.021$\pm$0.041 & 40.509 & 1.847$\pm$0.049 & 40.512 & 2.092$\pm$0.047 & 40.514 & 1.594$\pm$0.037 & 40.517 & 1.344$\pm$0.028 & 40.506 & 1.362$\pm$0.033 \\ 
40.511 & 0.628$\pm$0.022 & 40.503 & 2.064$\pm$0.049 & 42.305 & 1.982$\pm$0.052 & 42.308 & 2.165$\pm$0.049 & 42.311 & 1.633$\pm$0.038 & 42.314 & 1.366$\pm$0.028 & 42.302 & 1.391$\pm$0.033 \\ 
42.306 & 0.870$\pm$0.024 & 42.299 & 2.173$\pm$0.051 & 42.830 & 1.971$\pm$0.051 & 42.833 & 2.197$\pm$0.049 & 42.836 & 1.626$\pm$0.038 & 42.839 & 1.381$\pm$0.029 & 42.827 & 1.390$\pm$0.033 \\ 
42.831 & 0.863$\pm$0.024 & 42.824 & 2.204$\pm$0.052 & 45.297 & 1.846$\pm$0.048 & 45.300 & 2.150$\pm$0.048 & 45.303 & 1.619$\pm$0.038 & 45.306 & 1.402$\pm$0.029 & 45.294 & 1.372$\pm$0.033 \\ 
43.687 & 1.061$\pm$0.146 & 43.688 & 2.107$\pm$0.050 & 45.557 & 1.943$\pm$0.051 & 45.559 & 2.192$\pm$0.050 & 45.562 & 1.655$\pm$0.039 & 45.565 & 1.394$\pm$0.029 & 45.554 & 1.388$\pm$0.034 \\ 
45.298 & 0.837$\pm$0.023 & 45.291 & 2.138$\pm$0.050 & 48.750 & 2.109$\pm$0.055 & 48.752 & 2.345$\pm$0.053 & 48.755 & 1.738$\pm$0.040 & 48.758 & 1.422$\pm$0.030 & 48.747 & 1.398$\pm$0.034 \\ 
45.558 & 0.852$\pm$0.025 & 45.551 & 2.276$\pm$0.054 & 51.276 & 2.096$\pm$0.036 & 51.218 & 2.375$\pm$0.053 & 51.221 & 1.786$\pm$0.041 & 51.224 & 1.443$\pm$0.030 & 51.212 & 1.441$\pm$0.034 \\ 
48.751 & 1.188$\pm$0.029 & 48.744 & 2.481$\pm$0.058 & 51.606 & 2.206$\pm$0.056 & 51.609 & 2.359$\pm$0.053 & 51.612 & 1.763$\pm$0.041 & 51.615 & 1.413$\pm$0.029 & 51.603 & 1.418$\pm$0.034 \\ 
49.873 & 0.777$\pm$0.135 & 51.208 & 2.435$\pm$0.057 & 52.839 & 2.081$\pm$0.046 & 52.841 & 2.326$\pm$0.042 & 52.842 & 1.826$\pm$0.032 & 52.843 & 1.490$\pm$0.024 & 52.838 & 1.439$\pm$0.027 \\ 
51.216 & 1.076$\pm$0.027 & 51.600 & 2.478$\pm$0.058 & 53.867 & 2.078$\pm$0.053 & 53.870 & 2.284$\pm$0.051 & 53.873 & 1.751$\pm$0.040 & 53.876 & 1.427$\pm$0.029 & 53.864 & 1.403$\pm$0.034 \\ 
51.305 & 1.189$\pm$0.041 & 52.837 & 2.454$\pm$0.047 & 54.937 & 2.030$\pm$0.053 & 54.940 & 2.363$\pm$0.053 & 54.943 & 1.747$\pm$0.041 & 54.945 & 1.436$\pm$0.030 & 54.934 & 1.402$\pm$0.034 \\ 
51.607 & 1.165$\pm$0.028 & 53.861 & 2.378$\pm$0.055 & 55.927 & 2.097$\pm$0.054 & 55.930 & 2.266$\pm$0.051 & 55.933 & 1.740$\pm$0.040 & 55.936 & 1.424$\pm$0.029 & 55.924 & 1.408$\pm$0.034 \\ 
52.803 & 1.040$\pm$0.042 & 54.931 & 2.426$\pm$0.057 & 56.592 & 1.998$\pm$0.052 & 56.596 & 2.249$\pm$0.050 & 56.599 & 1.748$\pm$0.040 & 56.601 & 1.410$\pm$0.029 & 56.589 & 1.375$\pm$0.033 \\ 
53.868 & 0.963$\pm$0.025 & 55.921 & 2.421$\pm$0.056 & 58.063 & 2.012$\pm$0.052 & 58.066 & 2.271$\pm$0.051 & 58.069 & 1.732$\pm$0.040 & 58.072 & 1.414$\pm$0.029 & 58.060 & 1.432$\pm$0.034 \\ 
54.938 & 1.030$\pm$0.027 & 56.586 & 2.416$\pm$0.056 & 59.124 & 2.043$\pm$0.053 & 59.127 & 2.343$\pm$0.052 & 59.130 & 1.751$\pm$0.041 & 59.133 & 1.426$\pm$0.029 & 59.120 & 1.422$\pm$0.034 \\ 
55.929 & 0.973$\pm$0.026 & 58.057 & 2.362$\pm$0.055 & 59.585 & 2.237$\pm$0.047 & 59.587 & 2.390$\pm$0.042 & 59.588 & 1.758$\pm$0.031 & 59.590 & 1.428$\pm$0.023 & 59.517 & 1.448$\pm$0.038 \\ 
56.594 & 0.828$\pm$0.024 & 59.117 & 2.452$\pm$0.057 & 60.517 & 2.138$\pm$0.066 & 60.518 & 2.336$\pm$0.060 & 60.519 & 1.780$\pm$0.045 & 60.520 & 1.469$\pm$0.035 & 60.515 & 1.463$\pm$0.040 \\ 
58.064 & 0.909$\pm$0.024 & 59.583 & 2.361$\pm$0.045 & 60.854 & 2.249$\pm$0.066 & 60.855 & 2.416$\pm$0.060 & 60.857 & 1.775$\pm$0.044 & 60.858 & 1.460$\pm$0.033 & 60.852 & 1.430$\pm$0.038 \\ 
59.125 & 1.117$\pm$0.027 & 60.514 & 2.478$\pm$0.068 & 61.579 & 2.185$\pm$0.056 & 61.582 & 2.393$\pm$0.054 & 61.585 & 1.826$\pm$0.042 & 61.588 & 1.487$\pm$0.031 & 61.576 & 1.477$\pm$0.035 \\ 
59.586 & 1.144$\pm$0.029 & 60.851 & 2.499$\pm$0.066 & 62.610 & 2.063$\pm$0.046 & 62.611 & 2.293$\pm$0.042 & 62.612 & 1.777$\pm$0.032 & 62.614 & 1.444$\pm$0.024 & 62.608 & 1.446$\pm$0.028 \\ 
60.686 & 1.077$\pm$0.028 & 61.573 & 2.643$\pm$0.061 & 64.637 & 2.031$\pm$0.053 & 63.577 & 2.229$\pm$0.050 & 63.579 & 1.698$\pm$0.040 & 63.582 & 1.437$\pm$0.030 & 64.634 & 1.415$\pm$0.034 \\ 
61.580 & 0.831$\pm$0.024 & 62.607 & 2.326$\pm$0.046 & 67.042 & 2.041$\pm$0.053 & 64.640 & 2.282$\pm$0.051 & 64.643 & 1.700$\pm$0.039 & 64.646 & 1.416$\pm$0.029 & 67.039 & 1.423$\pm$0.034 \\ 
62.578 & 0.683$\pm$0.036 & 64.631 & 2.303$\pm$0.054 & 68.103 & 2.088$\pm$0.054 & 67.045 & 2.286$\pm$0.051 & 67.048 & 1.719$\pm$0.040 & 67.050 & 1.438$\pm$0.030 & 68.100 & 1.423$\pm$0.034 \\ 
63.575 & 0.765$\pm$0.031 & 67.036 & 2.294$\pm$0.054 & 69.100 & 2.175$\pm$0.056 & 68.106 & 2.335$\pm$0.052 & 68.108 & 1.711$\pm$0.040 & 68.111 & 1.418$\pm$0.029 & 69.097 & 1.418$\pm$0.034 \\ 
64.639 & 0.907$\pm$0.024 & 68.097 & 2.350$\pm$0.056 & 70.097 & 2.216$\pm$0.057 & 69.102 & 2.387$\pm$0.054 & 69.105 & 1.772$\pm$0.041 & 69.108 & 1.456$\pm$0.030 & 70.094 & 1.431$\pm$0.035 \\ 
65.695 & 0.300$\pm$0.100 & 69.094 & 2.506$\pm$0.059 & 72.054 & 2.262$\pm$0.044 & 70.100 & 2.449$\pm$0.055 & 70.102 & 1.785$\pm$0.042 & 70.105 & 1.456$\pm$0.030 & 72.052 & 1.459$\pm$0.026 \\ 
67.043 & 0.753$\pm$0.023 & 70.091 & 2.682$\pm$0.062 & 73.088 & 2.367$\pm$0.060 & 72.055 & 2.455$\pm$0.041 & 72.057 & 1.836$\pm$0.031 & 72.059 & 1.494$\pm$0.023 & 73.085 & 1.423$\pm$0.034 \\ 
68.104 & 0.819$\pm$0.024 & 72.050 & 2.600$\pm$0.046 & 76.512 & 2.415$\pm$0.049 & 73.091 & 2.494$\pm$0.056 & 73.094 & 1.877$\pm$0.044 & 73.097 & 1.476$\pm$0.031 & 76.510 & 1.451$\pm$0.027 \\ 
69.101 & 1.065$\pm$0.027 & 73.082 & 2.701$\pm$0.063 & 76.870 & 2.298$\pm$0.065 & 76.040 & 2.546$\pm$0.043 & 76.516 & 1.927$\pm$0.034 & 76.518 & 1.501$\pm$0.024 & 76.868 & 1.448$\pm$0.037 \\ 
70.098 & 0.848$\pm$0.025 & 76.508 & 2.773$\pm$0.051 & 77.802 & 2.412$\pm$0.062 & 76.514 & 2.496$\pm$0.044 & 76.873 & 1.865$\pm$0.045 & 76.875 & 1.494$\pm$0.033 & 77.799 & 1.481$\pm$0.036 \\ 
71.017 & 1.493$\pm$0.233 & 76.866 & 2.769$\pm$0.070 & 79.032 & 2.477$\pm$0.048 & 76.872 & 2.524$\pm$0.061 & 77.807 & 1.922$\pm$0.045 & 77.810 & 1.498$\pm$0.031 & 79.030 & 1.488$\pm$0.027 \\ 
72.054 & 0.557$\pm$0.017 & 77.797 & 2.800$\pm$0.065 & 79.531 & 2.548$\pm$0.064 & 77.805 & 2.634$\pm$0.059 & 79.036 & 1.949$\pm$0.033 & 79.038 & 1.518$\pm$0.024 & 79.528 & 1.476$\pm$0.035 \\ 
73.089 & 1.146$\pm$0.028 & 79.028 & 2.987$\pm$0.052 & 80.664 & 2.407$\pm$0.061 & 79.034 & 2.698$\pm$0.045 & 79.537 & 1.999$\pm$0.046 & 79.540 & 1.540$\pm$0.032 & 80.661 & 1.491$\pm$0.036 \\ 
76.042 & 0.740$\pm$0.028 & 79.525 & 2.983$\pm$0.068 & 81.525 & 2.398$\pm$0.061 & 79.534 & 2.688$\pm$0.059 & 80.669 & 1.925$\pm$0.045 & 80.672 & 1.537$\pm$0.032 & 81.522 & 1.452$\pm$0.035 \\ 
76.477 & 0.940$\pm$0.025 & 80.658 & 2.809$\pm$0.065 & 83.319 & 2.334$\pm$0.060 & 80.666 & 2.539$\pm$0.057 & 81.530 & 1.970$\pm$0.046 & 81.533 & 1.477$\pm$0.031 & 83.316 & 1.478$\pm$0.036 \\ 
76.711 & 1.043$\pm$0.027 & 81.519 & 2.772$\pm$0.064 & 84.453 & 2.413$\pm$0.061 & 81.527 & 2.587$\pm$0.058 & 83.325 & 1.870$\pm$0.044 & 83.327 & 1.520$\pm$0.032 & 84.450 & 1.484$\pm$0.036 \\ 
77.803 & 0.619$\pm$0.022 & 83.314 & 2.684$\pm$0.063 & 85.656 & 2.344$\pm$0.064 & 83.322 & 2.488$\pm$0.056 & 84.459 & 1.936$\pm$0.045 & 85.663 & 1.483$\pm$0.031 & 85.543 & 1.436$\pm$0.028 \\ 
79.033 & 1.159$\pm$0.025 & 84.447 & 2.832$\pm$0.065 & 86.521 & 2.373$\pm$0.060 & 84.456 & 2.599$\pm$0.058 & 85.660 & 1.941$\pm$0.046 & 87.650 & 1.508$\pm$0.032 & 86.518 & 1.482$\pm$0.035 \\ 
79.533 & 0.523$\pm$0.019 & 85.541 & 2.710$\pm$0.039 & 87.641 & 2.468$\pm$0.063 & 85.658 & 2.594$\pm$0.061 & 86.527 & 1.899$\pm$0.044 & 89.119 & 1.517$\pm$0.031 & 87.638 & 1.458$\pm$0.035 \\ 
80.665 & 0.806$\pm$0.024 & 86.515 & 2.870$\pm$0.065 & 89.110 & 2.471$\pm$0.062 & 86.524 & 2.594$\pm$0.057 & 87.647 & 1.937$\pm$0.045 & 92.377 & 1.513$\pm$0.032 & 89.107 & 1.466$\pm$0.035 \\ 
81.526 & 1.015$\pm$0.027 & 87.636 & 2.996$\pm$0.069 & 92.371 & 2.618$\pm$0.069 & 87.644 & 2.653$\pm$0.059 & 89.116 & 1.965$\pm$0.045 &     &    & 92.369 & 1.464$\pm$0.036 \\ 
83.320 & 0.727$\pm$0.023 & 89.104 & 2.972$\pm$0.067 &     &    & 89.113 & 2.656$\pm$0.058 & 92.375 & 1.992$\pm$0.047 &     &    &     &    \\ 
84.453 & 1.222$\pm$0.032 & 92.366 & 2.985$\pm$0.071 &     &    & 92.373 & 2.719$\pm$0.062 &     &    &     &    &     &    \\ 
85.447 & 1.097$\pm$0.046 &     &    &     &    &     &    &     &    &     &    &     &    \\ 
85.524 & 1.456$\pm$0.230 &     &    &     &    &     &    &     &    &     &    &     &    \\ 
85.657 & 1.143$\pm$0.033 &     &    &     &    &     &    &     &    &     &    &     &    \\ 
86.521 & 1.087$\pm$0.029 &     &    &     &    &     &    &     &    &     &    &     &    \\ 
87.642 & 1.121$\pm$0.029 &     &    &     &    &     &    &     &    &     &    &     &    \\ 
89.111 & 0.887$\pm$0.024 &     &    &     &    &     &    &     &    &     &    &     &    \\ 
\hline
  \end{tabular}
 \end{table}
\end{landscape}

\newpage

\begin{landscape}
 \begin{table}
  \caption{Wise and mm-wave light curves. 
  Flux density $F_\lambda$ in the optical is given in $10^{-14}$erg\,s$^{-1}$\,cm$^{-2}$\AA$^{-1}$.
Flux density $F_\nu$ in the mm is given in mJy. JD is listed as JD-2457300.}
  \label{tab:lightcurves2}
\scriptsize
  \begin{tabular}{cccccccccccc}
    \hline
    \multicolumn{6}{c}{Optical} & \multicolumn{4}{c}{mm} \\
    \multicolumn{2}{c}{V} & \multicolumn{2}{c}{R} & \multicolumn{2}{c}{I} & \multicolumn{2}{c}{E090} & \multicolumn{2}{c}{E150}  \\
    JD & $F_\lambda$ & JD & $F_\lambda$ & JD & $F_\lambda$ & JD & $F_\nu$ & JD & $F_\nu$  \\
\hline
13.429 & 1.360$\pm$0.020 & 13.424 & 2.187$\pm$0.036 & 13.433 & 1.884$\pm$0.034 & 41.785 &  8.24$\pm$0.81 & 41.785 &  9.69$\pm$0.97 \\ 
16.203 & 1.352$\pm$0.020 & 16.198 & 2.111$\pm$0.061 & 16.207 & 1.839$\pm$0.043 & 42.785 & 12.01$\pm$0.67 & 42.785 &  9.31$\pm$0.71 \\ 
17.200 & 1.402$\pm$0.021 & 17.213 & 2.231$\pm$0.071 & 17.204 & 1.872$\pm$0.046 & 43.762 & 11.67$\pm$0.68 & 43.762 &  8.83$\pm$0.63 \\ 
18.181 & 1.303$\pm$0.019 & 18.177 & 1.888$\pm$0.078 & 18.185 & 1.646$\pm$0.040 & 44.777 & 10.65$\pm$0.98 & 44.777 &  9.74$\pm$0.89 \\ 
19.179 & 1.294$\pm$0.021 & 19.174 & 1.978$\pm$0.048 & 19.183 & 1.681$\pm$0.040 & 45.820 & 12.09$\pm$0.66 & 45.820 & 10.92$\pm$0.75 \\ 
20.188 & 1.311$\pm$0.016 & 20.183 & 2.284$\pm$0.070 & 20.192 & 1.793$\pm$0.042 & 46.781 & 11.66$\pm$1.01 & 46.781 & 10.15$\pm$0.98 \\ 
23.183 & 1.331$\pm$0.018 & 23.179 & 2.000$\pm$0.047 & 23.188 & 1.775$\pm$0.034 & 48.797 &  6.30$\pm$1.00 & 48.797 &  6.47$\pm$1.11 \\ 
24.178 & 1.351$\pm$0.022 & 24.173 & 2.076$\pm$0.067 & 24.182 & 1.811$\pm$0.043 & 49.781 & 10.83$\pm$0.78 & 49.781 &  7.23$\pm$0.72 \\ 
25.177 & 1.367$\pm$0.020 & 25.172 & 2.075$\pm$0.061 & 25.181 & 1.861$\pm$0.046 & 58.770 & 13.73$\pm$1.12 & 58.770 &  9.71$\pm$1.05 \\ 
26.176 & 1.375$\pm$0.019 & 26.172 & 2.132$\pm$0.053 & 26.181 & 1.861$\pm$0.044 & 59.777 & 13.95$\pm$1.00 & 59.777 & 10.46$\pm$0.80 \\ 
27.176 & 1.399$\pm$0.018 & 28.171 & 2.159$\pm$0.053 & 27.181 & 1.906$\pm$0.038 & 60.777 & 12.64$\pm$0.78 & 60.777 &  9.39$\pm$0.80 \\ 
28.175 & 1.386$\pm$0.021 & 29.173 & 2.147$\pm$0.065 & 28.179 & 1.877$\pm$0.052 & 61.719 & 10.97$\pm$0.92 & 61.719 &  8.73$\pm$0.71 \\ 
35.207 & 1.383$\pm$0.009 & 35.202 & 2.173$\pm$0.041 & 29.182 & 1.869$\pm$0.041 & 62.703 & 12.53$\pm$1.09 & 62.703 &  9.72$\pm$0.85 \\ 
36.365 & 1.394$\pm$0.015 & 36.361 & 2.182$\pm$0.044 & 35.211 & 1.916$\pm$0.030 & 64.770 & 11.71$\pm$0.82 & 64.770 & 11.70$\pm$1.27 \\ 
38.169 & 1.384$\pm$0.017 & 37.434 & 2.132$\pm$0.057 & 36.370 & 1.904$\pm$0.032 & 65.777 &  6.85$\pm$2.04 & 65.777 &  6.59$\pm$3.40 \\ 
39.168 & 1.396$\pm$0.021 & 38.165 & 2.191$\pm$0.062 & 38.174 & 1.884$\pm$0.042 & 66.758 & 10.51$\pm$1.49 & 66.758 & 11.20$\pm$2.37 \\ 
40.199 & 1.419$\pm$0.016 & 39.163 & 2.297$\pm$0.087 & 39.172 & 1.888$\pm$0.041 & 67.820 & 11.28$\pm$0.79 & 67.820 &  8.72$\pm$0.93 \\ 
43.177 & 1.390$\pm$0.014 & 40.194 & 2.270$\pm$0.052 & 40.203 & 1.951$\pm$0.036 & 68.820 & 10.65$\pm$0.78 & 68.820 & 10.53$\pm$0.86 \\ 
44.179 & 1.384$\pm$0.018 & 43.173 & 2.198$\pm$0.039 & 43.181 & 1.875$\pm$0.037 & 69.828 & 10.51$\pm$1.39 & 69.828 &  4.43$\pm$1.93 \\ 
46.170 & 1.387$\pm$0.018 & 44.175 & 2.140$\pm$0.097 & 44.174 & 1.856$\pm$0.038 & 70.781 & 12.41$\pm$1.98 & 70.781 & 13.02$\pm$4.26 \\ 
47.172 & 1.354$\pm$0.015 & 46.166 & 2.180$\pm$0.053 & 46.174 & 1.858$\pm$0.040 & 71.785 & 11.40$\pm$0.85 & 71.785 & 10.83$\pm$0.82 \\ 
49.176 & 1.387$\pm$0.011 & 47.174 & 2.093$\pm$0.104 & 47.183 & 1.758$\pm$0.052 & 72.738 & 11.74$\pm$1.04 & 72.738 & 10.46$\pm$0.92 \\ 
50.197 & 1.383$\pm$0.011 & 49.171 & 2.151$\pm$0.046 & 49.180 & 1.856$\pm$0.038 & 73.734 & 12.75$\pm$1.01 & 73.734 &  8.77$\pm$1.04 \\ 
51.333 & 1.389$\pm$0.014 & 50.192 & 2.147$\pm$0.034 & 50.201 & 1.873$\pm$0.024 & 74.699 & 14.12$\pm$1.33 & 74.699 & 13.35$\pm$1.57 \\ 
52.166 & 1.363$\pm$0.015 & 51.329 & 2.174$\pm$0.036 & 51.337 & 1.847$\pm$0.036 & 75.695 & 11.45$\pm$1.11 & 75.695 & 10.21$\pm$1.00 \\ 
53.174 & 1.366$\pm$0.015 & 52.162 & 2.016$\pm$0.041 & 52.171 & 1.761$\pm$0.033 & 76.699 & 10.94$\pm$0.74 & 76.699 & 11.21$\pm$0.99 \\ 
54.165 & 1.412$\pm$0.017 & 53.170 & 2.050$\pm$0.045 & 53.179 & 1.790$\pm$0.029 & 77.699 & 13.10$\pm$1.13 & 77.699 &  9.09$\pm$0.95 \\ 
55.168 & 1.397$\pm$0.018 & 54.161 & 2.231$\pm$0.055 & 54.170 & 1.916$\pm$0.037 & 78.773 & 10.08$\pm$0.96 & 78.773 &  9.27$\pm$0.74 \\ 
56.170 & 1.420$\pm$0.016 & 55.164 & 2.135$\pm$0.045 & 55.173 & 1.909$\pm$0.038 & 79.766 &  7.90$\pm$0.97 & 79.766 &  6.92$\pm$1.34 \\ 
57.191 & 1.352$\pm$0.016 & 56.166 & 2.231$\pm$0.057 & 56.174 & 1.929$\pm$0.039 & 80.734 & 10.41$\pm$1.06 & 80.734 & 10.00$\pm$0.80 \\ 
59.180 & 1.423$\pm$0.019 & 57.174 & 2.104$\pm$0.093 & 57.195 & 1.739$\pm$0.040 & 81.738 & 10.05$\pm$1.02 & 81.738 & 10.73$\pm$0.93 \\ 
61.208 & 1.429$\pm$0.013 & 59.176 & 2.249$\pm$0.061 & 59.184 & 1.912$\pm$0.039 & 82.734 & 11.77$\pm$0.86 & 82.734 & 12.56$\pm$0.79 \\ 
62.209 & 1.429$\pm$0.009 & 61.200 & 2.259$\pm$0.032 & 61.209 & 1.951$\pm$0.030 & 83.750 &  8.48$\pm$1.25 & 83.750 &  8.10$\pm$1.28 \\ 
63.329 & 1.417$\pm$0.020 & 62.205 & 2.214$\pm$0.029 & 62.214 & 1.953$\pm$0.025 & 84.742 & 11.61$\pm$1.16 & 84.742 & 11.32$\pm$1.49 \\ 
64.276 & 1.414$\pm$0.020 & 63.325 & 2.332$\pm$0.058 & 63.333 & 1.926$\pm$0.045 & 85.746 & 10.76$\pm$0.81 & 85.746 &  9.58$\pm$0.91 \\ 
65.177 & 1.408$\pm$0.018 & 64.272 & 2.197$\pm$0.045 & 64.280 & 1.891$\pm$0.041 & 86.715 & 11.06$\pm$0.64 & 86.715 & 10.93$\pm$0.69 \\ 
66.169 & 1.416$\pm$0.015 & 65.172 & 2.223$\pm$0.056 & 65.181 & 1.883$\pm$0.040 &     &     &     &     \\ 
67.168 & 1.411$\pm$0.021 & 66.165 & 2.212$\pm$0.046 & 66.174 & 1.916$\pm$0.035 &     &     &     &     \\ 
68.176 & 1.439$\pm$0.016 & 67.163 & 2.204$\pm$0.047 & 67.172 & 1.900$\pm$0.042 &     &     &     &     \\ 
71.218 & 1.421$\pm$0.012 & 68.172 & 2.275$\pm$0.045 & 68.181 & 1.927$\pm$0.042 &     &     &     &     \\ 
76.168 & 1.395$\pm$0.014 & 71.211 & 2.252$\pm$0.028 & 71.220 & 1.927$\pm$0.027 &     &     &     &     \\ 
77.169 & 1.409$\pm$0.016 & 76.164 & 2.230$\pm$0.035 & 76.173 & 1.871$\pm$0.034 &     &     &     &     \\ 
78.172 & 1.406$\pm$0.016 & 77.165 & 2.200$\pm$0.043 & 77.173 & 1.864$\pm$0.037 &     &     &     &     \\ 
79.168 & 1.420$\pm$0.016 & 78.168 & 2.178$\pm$0.045 & 78.177 & 1.862$\pm$0.038 &     &     &     &     \\ 
80.204 & 1.408$\pm$0.019 & 79.164 & 2.224$\pm$0.050 & 79.173 & 1.889$\pm$0.041 &     &     &     &     \\ 
81.174 & 1.391$\pm$0.018 & 80.200 & 2.211$\pm$0.036 & 80.208 & 1.892$\pm$0.037 &     &     &     &     \\ 
83.172 & 1.415$\pm$0.019 & 81.169 & 2.179$\pm$0.050 & 81.178 & 1.878$\pm$0.043 &     &     &     &     \\ 
84.172 & 1.382$\pm$0.016 & 83.168 & 2.172$\pm$0.046 & 83.176 & 1.902$\pm$0.039 &     &     &     &     \\ 
85.172 & 1.313$\pm$0.060 & 84.167 & 2.171$\pm$0.040 & 84.176 & 1.775$\pm$0.035 &     &     &     &     \\ 
86.173 & 1.429$\pm$0.019 & 85.168 & 2.180$\pm$0.066 & 85.177 & 1.740$\pm$0.127 &     &     &     &     \\ 
92.180 & 1.450$\pm$0.023 & 86.169 & 2.264$\pm$0.051 & 86.177 & 1.915$\pm$0.050 &     &     &     &     \\ 
    &    & 92.176 & 2.305$\pm$0.060 & 92.185 & 1.906$\pm$0.042 &     &     &     &     \\ 
\hline
  \end{tabular}
 \end{table}
\end{landscape}

\newpage

\begin{landscape}
 \begin{table}
  \caption{\xmm /OM light curves. Flux density $F_\lambda$ is given in $10^{-14}$erg\,s$^{-1}$cm$^{-2}$\AA$^{-1}$.   
JD is listed as JD-2457300.}
  \label{tab:lightcurves3}
  \begin{tabular}{ccccccc}
    \hline
    \multicolumn{7}{c}{\xmm /OM} \\
    JD & {w2} & {m2} & {w1} & {U} & {B} & {V}  \\
      & $F_\lambda$ & $F_\lambda$ & $F_\lambda$ & $F_\lambda$ & $F_\lambda$ & $F_\lambda$  \\
\hline
51.527 & 2.490$\pm$0.065 & 2.170$\pm$0.076 & 2.360$\pm$0.044 & 1.730$\pm$0.038 & 1.410$\pm$0.023 & 1.390$\pm$0.033 \\ 
72.084 & 2.660$\pm$0.070 & 2.290$\pm$0.080 & 2.450$\pm$0.046 & 1.860$\pm$0.041 & 1.440$\pm$0.024 & 1.430$\pm$0.034 \\ 
80.030 & 2.750$\pm$0.073 & 2.450$\pm$0.086 & 2.610$\pm$0.049 & 1.990$\pm$0.043 & 1.510$\pm$0.025 & 1.480$\pm$0.035 \\ 
82.001 & 2.750$\pm$0.073 & 2.370$\pm$0.083 & 2.560$\pm$0.048 & 1.950$\pm$0.042 & 1.510$\pm$0.025 & 1.480$\pm$0.035 \\ 
83.966 & 2.720$\pm$0.072 & 2.410$\pm$0.084 & 2.570$\pm$0.048 & 1.940$\pm$0.042 & 1.490$\pm$0.025 & 1.450$\pm$0.034 \\ 
85.950 & 2.760$\pm$0.073 & 2.460$\pm$0.086 & 2.600$\pm$0.049 & 1.980$\pm$0.043 & 1.520$\pm$0.025 & 1.470$\pm$0.035 \\ 
\hline
  \end{tabular}
 \end{table}
\end{landscape}

\end{document}